\documentclass[preprint,showpacs,preprintnumbers,amsmath,amssymb]{revtex4}
\usepackage{graphicx}
\usepackage{dcolumn}
\usepackage{bm}
\usepackage{amssymb}
\usepackage{amsmath}
\usepackage{latexsym}

\begin{document}

\title{Origin of the orbital period change in  contact binary stars}

\author{V.V. Sargsyan$^{1,2}$, H. Lenske$^{2}$, G.G. Adamian$^{1}$, and N.V. Antonenko$^{1}$}

\affiliation{
$^{1}$Joint Institute for Nuclear Research, 141980 Dubna, Russia,\\
$^{2}$Institut f\"ur Theoretische Physik der
Justus--Liebig--Universit\"at, D--35392 Giessen, Germany}

 \date{\today}

\begin{abstract}
The evolution of  contact binary  star systems in mass asymmetry (transfer) coordinate is considered.
The orbital period changes is explained by an
evolution in mass asymmetry  towards the symmetry (symmetrization of binary system).
It is predicted that a decreasing and an increasing orbital periods
are related, respectively, with the non-overlapping and overlapping  stage of the binary star during its symmetrization.
A huge amount of energy $\Delta U\approx 10^{41}$ J is converted  from the
 the potential energy into internal energy of the stars during the symmetrization.
As shown, the merger of  stars in the binary systems, including KIC 9832227, is energetically an unfavorable process.
The sensitivity of the calculated results to the values of  total mass and orbital angular momentum
is analyzed.
\end{abstract}


\maketitle

\section{Introduction}\label{sec:Intro}

Overcontact, contact, and near-contact
binaries, forming di-star compounds
with the average distances between the stars  of the
same order as the sum of their radii \cite{Boya:2002,Eggleton:2006,Vasil:2012,Cher:2013} are of  great interest for stellar evolution.
The compact binary stars are a  good laboratory for a wide variety of astrophysical
phenomena, for example, the mass transfer between stars.
The information on  formation and evolution of compact binaries
is required   to understand the processes observed in isolated stars.
%
%
The observations of the stages of evolution
in these binaries  provide a verification of our
understanding of the inner structure and dynamical interaction of stars.

It has been found that the orbital period of W-type overcontact
binary GW Cep ($q=M_{1i}/M_{2i}=0.37$, $\eta_i=(M_{1i}-M_{2i})/(M_{1i}+M_{2i})=0.46$, $M_{1i}$ and $M_{2i}$ are the masses of stars in the binary)
is decreasing with time \cite{Qian}.
For the overcontact binaries VY Cet  ($q=0.67$, $\eta_i=0.20$) and V700 Cyg  ($q=0.65$, $\eta_i=0.21$),
a cyclic oscillations have been found to be superimposed on a secular period increase.
This effect has been explained either  by the strong external perturbation,
i.e. by a close-by third object, or by the magnetic activity cycles.
For the EM Lac  ($q=0.63$, $\eta_i=0.23$)  and AW Vir  ($q=0.76$, $\eta_i=0.14$)
binaries, the periods show a secular increase.
 It has been concluded that the period variations  of a W UMa-type binary star
 is correlated with the mass ratio $q$ and the mass $M_{1i}$  of the primary component \cite{Qian}.
The low mass ratio $q$ in binaries usually results in a decreasing period, while
the periods of larger-$q$ systems are increasing.


Because mass transfer is an important observable for close
binary systems in which the two stars are nearly in contact
\cite{Boya:2002,Eggleton:2006,Vasil:2012,Cher:2013,Qian,IJMPE,IJMPE2},
it is   necessary to study the influence of evolution of these stellar
systems in the mass asymmetry coordinate
$\eta = (M_1-M_2)/M$ [$M_k$ ($k$ =1,2) are the stellar
masses and $M=M_1+M_2$] on the orbital period variations. This is the main aim of the present article.
As in our previous works \cite{IJMPE,IJMPE2}, we  analyze the
total potential energy $U(\eta)$ and the orbital period as
functions of $\eta$ at fixed total mass $M=M_1+M_2$
and orbital angular momentum $L=L_i$ of the system.

%

\section{Theoretical Method}\label{sec:Theory}
%

The total potential energy of the di-star system
\begin{eqnarray}
U=U_1+U_2+V
\label{eq_pot}
\end{eqnarray}
is given by the sum of the potential energies $U_k$ ($k=1,2$) of  two stars and star-star interaction potential $V$.
The radiation energy is neglected because
 the absolute values of the gravitational energy and the intrinsic kinetic energy
 are much  larger than  the radiation energy.
The energy of the star  "$k$" is
\begin{eqnarray}
U_{k}=-\omega_k\frac{G M_k^2}{2R_k},
\label{eq_pot2}
\end{eqnarray}
where $G$, $M_k$, and $R_k$ are the gravitational constant, mass, and radius of the star, respectively.
Employing  the values of the dimensionless structural factor
\begin{eqnarray}
\omega_k=1.644\left(\frac{M_{\odot}}{M_k}\right)^{1/4}
\label{eq_omega2}
\end{eqnarray}
and radius
$$R_k=R_{\odot}\left(\frac{M_k}{M_{\odot}}\right)^{2/3}$$
of the star from  the model of Ref. \cite{Vasil:2012},
we obtain
\begin{eqnarray}
U_{k}&=&-\omega_0G M_{k}^{13/12}/2,\nonumber \\
\omega_0&=&1.644\frac{M_{\odot}^{11/12}}{R_{\odot}},
\label{eq_pot1}
\end{eqnarray}
where   $M_{\odot}$ and $R_{\odot}$ are   mass  and radius of the Sun, respectively.
Because the average density $\rho_k=M_{\odot}\rho_{\odot}/M_k$ ($\rho_{\odot}$
is the average density of the Sun) increase with decreasing the mass $M_k$ of star \cite{Vasil:2012},
the structural factor $\omega_k$ depends on $M_k$ in Eq. (\ref{eq_omega2}).
The change of $\eta$ from 0 to 1 leads to the change  of $\omega_1$ by about of 16\%.
Note that, in  general, the dimensionless structural factor $\omega_i$
is   determined by   the density profile  of the star.  In the present paper, we
employ    the values of the structural factor of the stars from  the model of Ref. \cite{Vasil:2012}.
This model  well describes the
observable temperature-radius-mass-luminosity relations of stars, especially binary stars,
the spectra of seismic oscillations of the Sun, distribution of stars on their masses, magnetic
fields of stars and etc.   The stellar radii, masses, and temperatures are expressed by the corresponding ratios of the fundamental
constants, and the individuality of stars is determined by two parameters - by the charge and mass numbers of nuclei,
from which the star is composed \cite{Vasil:2012}.

Because the two stars rotate with respect to each other around the common center of mass,
the star-star interaction potential contains, together with the gravitational energy
of  interaction $V_G$ of two stars,  the kinetic energy of orbital rotation $V_R$:
\begin{eqnarray}
V(R)=V_G+V_R=V_G+\frac{L^2}{2\mu R^2},
\label{eq_pot3}
\end{eqnarray}
where $L$ is the orbital angular momentum of the di-star which is conserved during the conservative mass transfer and
$\mu=\frac{M_1M_2}{M}$ is the reduced mass.
At $R\ge R_t=R_1+R_2$ and  $R\le R_t$,
\begin{eqnarray}
V_G(R)=-\frac{GM_1M_2}{R}
\label{eq_pot3n}
\end{eqnarray}
and
\begin{eqnarray}
V_G(R)=-\frac{GM_1M_2}{2R_t}\left[3-\frac{R^2}{R_t^2}\right],
\label{eq_pot3nn}
\end{eqnarray}
respectively \cite{NPA}.
Here, $R_t$ is the touching distance.
From the fix point conditions $\partial V/\partial R|_{R=R_m}=0$ and  $\partial^2 V/\partial R^2|_{R=R_m}>0$,
we find the relative equilibrium distance between two stars corresponding to the minimum of $V$:
\begin{eqnarray}
R_m=\frac{L^2}{G\mu^2 M}
\label{eq_pot4n}
\end{eqnarray}
at $R_m\ge R_t$
or
\begin{eqnarray}
R_m=\left(\frac{L^2R_t^3}{G\mu^2 M}\right)^{1/4}
\label{eq_pot4nn}
\end{eqnarray}
at $R_m\le R_t$.
Finally, one can derive the   expression for the star-star interaction potential
\begin{eqnarray}
V(R_m)=-\frac{GM_1M_2}{2R_m}
\label{eq_pot5nn}
\end{eqnarray}
at $R_m\ge R_t$
or
\begin{eqnarray}
V(R_m)=-\frac{GM_1M_2}{R_t}\left[\frac{3}{2}-\frac{R_m^2}{R_t^2}\right]
\label{eq_pot5nnn}
\end{eqnarray}
at $R_m\le R_t$.
In Eqs. (\ref{eq_pot5nn}) and (\ref{eq_pot5nnn}),   $R_m$ is
the semi-major axis of  the elliptical relative orbit.
Note that
for the di-star systems considered, $v(R_m)=(GM/R_m)^{1/2}\ll c$, where $v$ and $c$ is the velocities of orbital motion and light, respectively, and one can
neglect the relativistic effects. Since  $GM_k/R_m\ll c^2$, the gravitational field  can be considered weak.
Because of these facts, we use the Newtonian law of gravity.

Using the mass asymmetry coordinate $\eta$ instead of masses $M_1=\frac{M}{2}(1+\eta)$ and $M_2=\frac{M}{2}(1-\eta)$,
we rewrite   the final expression (\ref{eq_pot}) for the total potential energy as
\begin{eqnarray}
U=-\frac{GM_{\odot}^{2}}{2R_{\odot}}\left(\alpha[(1+\eta)^{13/12}+(1-\eta)^{13/12}]+\beta_1 [1-\eta^2]^{3}\right)
\label{eq_pot7}
\end{eqnarray}
at $R_m\ge R_t$
or
\begin{eqnarray}
U&=&-\frac{GM_{\odot}^{2}}{2R_{\odot}}\left(\alpha[(1+\eta)^{13/12}+(1-\eta)^{13/12}]\right.\nonumber\\
&+&\left.\beta_2\frac{1-\eta^2}{(1+\eta)^{2/3}+(1-\eta)^{2/3}}\left[\frac{3}{2}-\frac{\gamma}{[1-\eta^2][(1+\eta)^{2/3}+(1-\eta)^{2/3}]^{1/2}}\right]\right)
\label{eq_pot7}
\end{eqnarray}
at $R\le R_t$,
where
$$\alpha=1.644\left(\frac{M}{2M_{\odot}}\right)^{13/12},$$
$$\beta_1=\frac{GM^5}{64L^2}=\frac{GM_{\odot}^5}{2L^2}\left(\frac{M}{2M_{\odot}}\right)^{5},$$
$$\beta_2=2\left(\frac{M}{2M_{\odot}}\right)^{4/3},$$
and
$$\gamma=\frac{2^{7/3}LM_{\odot}^{1/3}}{\left(GR_{\odot}M^{11/3}\right)^{1/2}}.$$
Here, we assume that  the orbital angular momentum $L$ and the total mass $M$ are conserved during
the conservative evolution of the di-star in  the mass asymmetry coordinate $\eta$.
The orbital angular momentum $L$ is calculated by using the experimental masses $M_{k,i}$
of stars and  period $P_{orb,i}$ of their orbital rotation \cite{Qian,Egg}.
As seen from Eq. (\ref{eq_pot7}), the stability of the binary star system depends on
the orbital angular momentum $L$ the total mass $M$, and the structural factor.

To obtain the period  $P_{orb}=\frac{2\pi}{\omega_{orb}}=\frac{2\pi\mu R_m^2}{L}$
of orbital rotation with frequency $\omega_{orb}$, we use the relation between $L$ and $R_m$.
At $R_m>R_t$ and $R_m\le R_t$, we have the periods
\begin{eqnarray}
P^{>}_{orb}= 2\pi\left(\frac{ R_m^3}{G M }\right)^{1/2}
\label{eq_pot4nn}
\end{eqnarray}
and
\begin{eqnarray}
P^{<}_{orb}= 2\pi\left(\frac{ R_t^3}{G M }\right)^{1/2},
\label{eq_pot4nnn}
\end{eqnarray}
respectively.
As seen from formulas, at initial $|\eta_i|$ smaller than the position
$|\eta_b|$ of barrier of the potential energy $U$ (see Sect. III) and $R_m>R_t$ ($R_m\le R_t$),
the system moves towards the symmetric configuration and,
respectively, $\eta$ decreases, $R_m$ decreases ($R_t$ increases),
and, finally, the $P^{>}_{orb}$ decreases ($P^{<}_{orb}$ increases).


\section{Application to Close Binaries}\label{sec:Results}

Many  di-star systems have different $M$ and $L$ (Table I) \cite{Qian,Egg} and, correspondingly, the potential energy shapes.
The  potential energies (driving potentials) $U(\eta)$ of the close di-star systems
versus $\eta$  are presented in Figs.~1 and 2. In the calculations, we assume that the orbital angular momentum and the
total mass are conserved during the conservative evolution of the di-star in mass asymmetry coordinate $\eta$.
%
For all binary systems shown,  the potential energies have
the barriers at $\eta=\pm\eta_b$ and the minimum at $\eta=\eta_m=0$.
The barrier in $\eta$ appears as a result of the interplay
between the total gravitational energy $U_1+U_2$ of the stars and the star-star interaction potential $V$.
These values have different behavior as a function of mass asymmetry: $U_1+U_2$ decreases and $V$ increases
with changing $\eta$ from $0$ to $\pm 1$.
One should stress that the driving potentials $U(\eta)$ for the di-star systems looks
like the driving potentials for the microscopic dinuclear systems \cite{Adamian:2012,Adamian:2014}.
The collective coordinate $\eta$, treated a separate dynamical degree of freedom,  plays
a comparable important role   in macroscopic object as well
as in microscopic dinuclear systems.
Note that the same conclusion was drawn  in Refs. \cite{IJMPE,IJMPE2}.

The evolution of  di-star system   depends on the initial  mass asymmetry $\eta=\eta_i$ at its formation.
The original di-star is asymmetric and $|\eta_i|<|\eta_b|$, then it is energetically favorable to evolve
in  $\eta$ towards a configuration   at $\eta=0$, that is, to form a symmetric  di-star system.
The matter of a heavy star can move to an adjacent light star enforcing the symmetrization of di-star without additional driving energy.
The symmetrization  of asymmetric binary star leads to the decrease of potential energy $U$,  i. e.
the transformation of the potential energy into internal energy of the stars.
A huge amount of energy $\Delta U\approx 10^{41}$ J is released and converted during the symmetrization (see Figs. 1  and 2).
The resulting symmetric di-star is created at large excitation energy. So, the binary star systems are the sources of
energy in the Universe.
Note that for all binary stars considered the energy of one single star ($|\eta|=1$) is larger than the energy of the
symmetric binary ($\eta=0$) which means that  the merger of  stars in the binary system is energetically an unfavorable process.

A spectacular recent case is KIC 9832227 which
was predicted \cite{Molnar:2017} to be merge in 2022, enlightening the sky as a red nova.
For the  KIC 9832227 ($|\eta_i|=0.63$, $|\eta_b|=0.84$), we predict that a fast merger is excluded (see Fig. 2). This di-star
is driven instead towards the mass symmetry. It should be stressed that the observational data of Ref. \cite{Socia}
negate the merger of KIC 9832227 in 2022.

For the KIC 9832227, we studied the change of potential energy surface with the variations of the
total mass $M$ and orbital angular momentum $L$ of system (Figs. 2 and 3).
In this way, we take effectively into consideration  the losses of mass and angular momentum
of the binary star during its evolution.
The mass or angular momentum may be removed from the system via stellar wind or gravitational radiation.
As seen, a loss of the orbital angular momentum at a fixed mass $M$ increases the symmetrization rate.
The simultaneous losses of $M$ and $L$ ($\Delta M\sim\Delta L$) by 20\% or 50\% weakly influence the shape of potential energy.
The decrease of $M$ at fixed $L$ has a more pronounced effect  (the depth of the minimum in $U(\eta)$ decreases)
but the evolution to the global symmetric minimum is still energetically favored.
So, the realistic  simultaneous
losses of $M$ and $L$ almost do not influence the rate of symmetrization of the system.
Note that a loss of the orbital angular momentum at the fixed mass increase the rate   of symmetrization of the system.

For KIC 9832227 system  ($|\eta_i|=0.63$),
the mass is transferring from the heavy star to light companion, and the relative distance ($R_m>R_t=R_1+R_2$) between
two stars and the period $P^{>}_{orb}$
of the orbital rotation are decreasing. The evolution in $\eta$ leads to the touching configuration ($R_m=R_t=R_1+R_2$) of stars
at some critical mass asymmetry $|\eta|=|\eta_t|\approx 0.45$ (Fig. 4).
Further evolution in $\eta$ leads to a configuration with  partial overlap
($R_m<R_t$) of the stars. So, at $|\eta|\le|\eta_t|$ the period $P^{<}_{orb}$ of the orbital rotation is slightly increasing because $P^{<}_{orb}\sim R_t^{3/2}$ and
$R_t$ increases with decreasing $\eta$.
Thus, once the system has crossed the point $\eta=\eta_t$ and
 the phase of partial overlap is entered, the evolution of period  changes abruptly.
The other  contact binaries considered in Fig. 4 show a similar behavior of period.

For the
binary GW Cep ($|\eta_i|=0.46$), $|\eta_i|>|\eta_t|$, the system moves towards mass symmetry and the orbital period is decreasing with time.
%
For the almost symmetric EM Lac  ($|\eta_i|=0.23$)  and AW Vir  ($|\eta_i|=0.14$) binaries, $|\eta_i|<|\eta_t|$
and the periods show the secular increase.
So, one can conclude that the period variations  of a W UMa-type binary star
 is correlated with the mass asymmetry evolution towards the global symmetric minimum.
%
At low mass ratio $q$, i. e.  or the large mass asymmetry $\eta_i$,    binaries usually show a decreasing period because $|\eta_i|>|\eta_t|$
($R_m>R_t$), while
the periods in systems with high $q$ (or small $\eta_i$)  are increasing because $|\eta_i|<|\eta_t|$ ($R_m<R_t$).





\section{Summary}\label{sec:Summary}

For all  contact di-star systems considered,  the potential energies have
symmetric barriers at $\eta=\pm|\eta_b|$ and the minimum at $\eta=\eta_m=0$.
The di-star system is  initially  formed with $\eta=\eta_i$ and $0<|\eta_i|<|\eta_b|$.
The two stars start to exchange matter and the system
is driven to the symmetric di-star configuration
(towards a global minimum of the potential landscape).
The   mass asymmetry coordinate
is governing the  symmetrization driven by the mass transfer
process of two stars.  The losses of the total mass and orbital angular momentum
weakly influence the symmetrization of system.
Note that the merger of the binary star systems considered  is  energetically not favored.


The orbital period changes can be plausibly explained by an
evolution in mass asymmetry $\eta$ towards the symmetry. We predicted that a decreasing and an increasing orbital periods
are related, respectively, with the non-overlapping ($|\eta_i|>|\eta_t|$, $R_m > R_1+R_2$)
and overlapping ($|\eta_i|<|\eta_t|$, $R_m < R_1+R_2$) stage of the binary star during its symmetrization.
Thus, the observations of changing periods allows us to distinguish between these two stages of the binary star systems.

%

\section{Acknowledgements}
This work was partially supported by  Russian Foundation for Basic Research (Moscow), grant number 17-52-12015,  and
DFG (Bonn), contract Le439/16.


\begin{table}
\small\addtolength{\tabcolsep}{0pt}
\caption{
The observed data $M_1/M_{\odot}$,  $M_2/M_{\odot}$, and $P_{orb}$ are from Refs.
\protect\cite{Qian,Egg,Molnar:2017}.
}
\begin{tabular*}{\textwidth}{@{\extracolsep{\fill}}|c|c|c|c|}
\hline
 Di-star  & $\frac{M_1}{M_{\odot}}$ & $\frac{M_2}{M_{\odot}}$ &  $P_{orb}$     \\
          &                         &                         &    days         \\
\hline
AH Aur     & 1.68  & 0.28  &  0.4941   \\
AP Aur     & 2.05  & 0.50  &  0.5694   \\
DN Aur     & 1.44  & 0.30  &  0.6169   \\
AW Vir     & 1.11  & 0.84  &  0.3540   \\
AW UMa     & 1.38  & 0.14  &  0.4387   \\
HV UMa     & 2.84  & 0.54  &  0.7108   \\
KIC 9832227& 1.40  & 0.32  &  0.4583   \\
HV Aqr     & 1.31  & 0.19  &  0.3734  \\
GX And     & 1.23  & 0.29  &  0.4122   \\
RR Cen     & 2.09  & 0.45  &  0.6060    \\
EM Lac     & 1.06  & 0.67  &  0.3891    \\
GW Cep     & 1.06  & 0.39  &  0.3188    \\
V700  Cyg  & 0.92  & 0.60  &  0.3400   \\
V870  Ara  & 1.34  & 0.11  &  0.3997    \\

\hline
\end{tabular*}
\label{tab1}
\end{table}

\begin{figure}[ht]
   \resizebox{\hsize}{!}
{\includegraphics[width=0.45\linewidth]{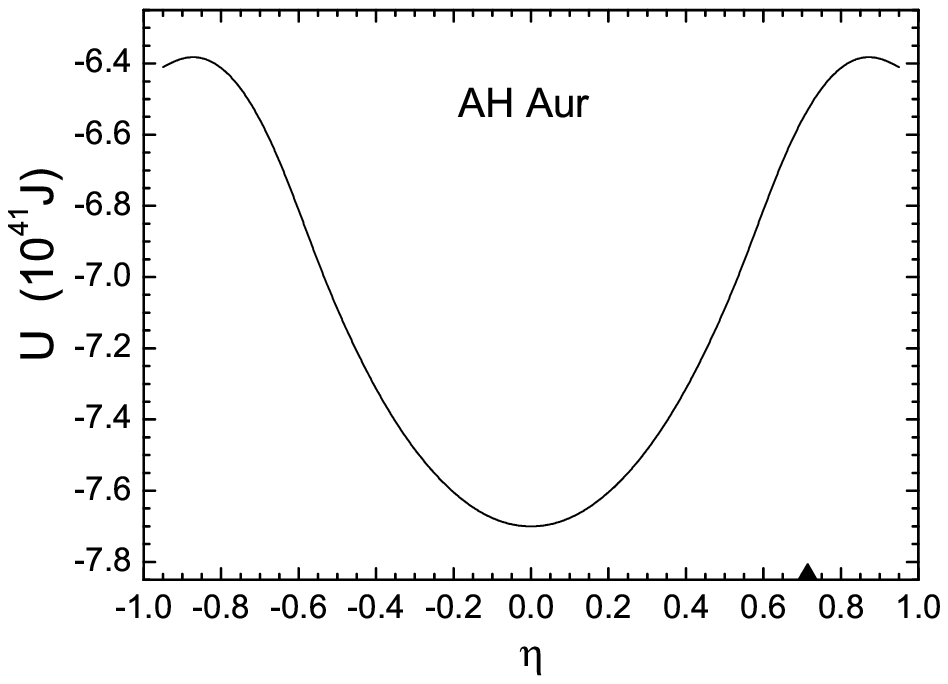}
\includegraphics[width=0.45\linewidth]{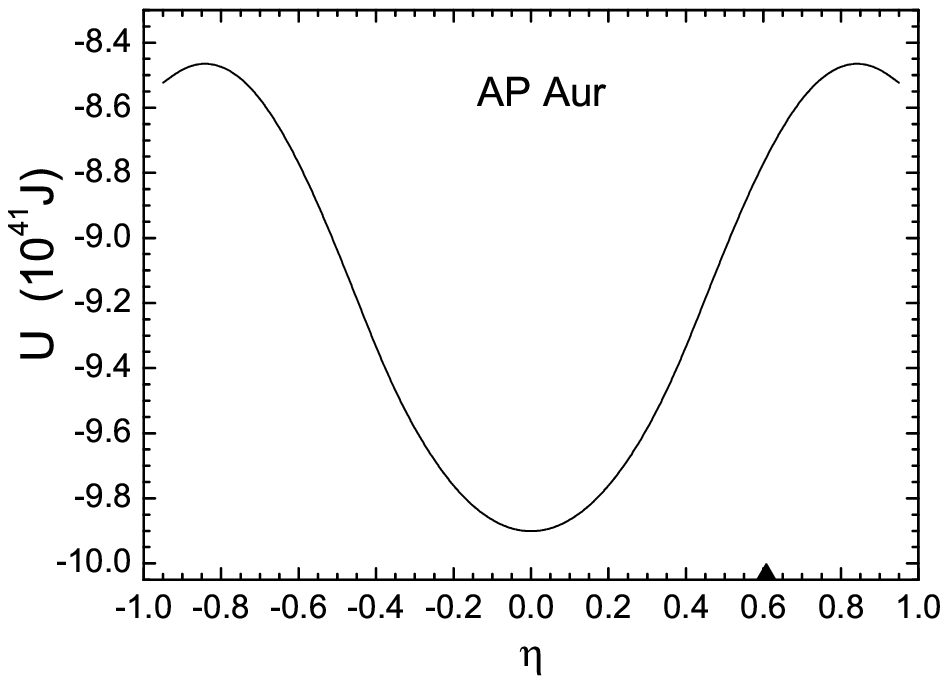}}
{\includegraphics[width=0.5\linewidth]{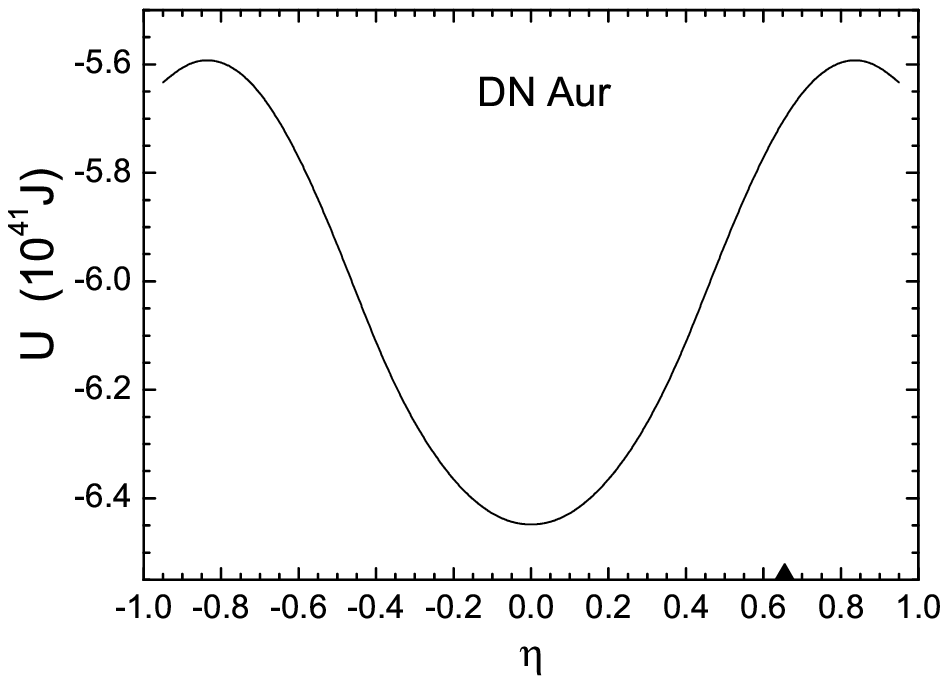}
\includegraphics[width=0.49\linewidth]{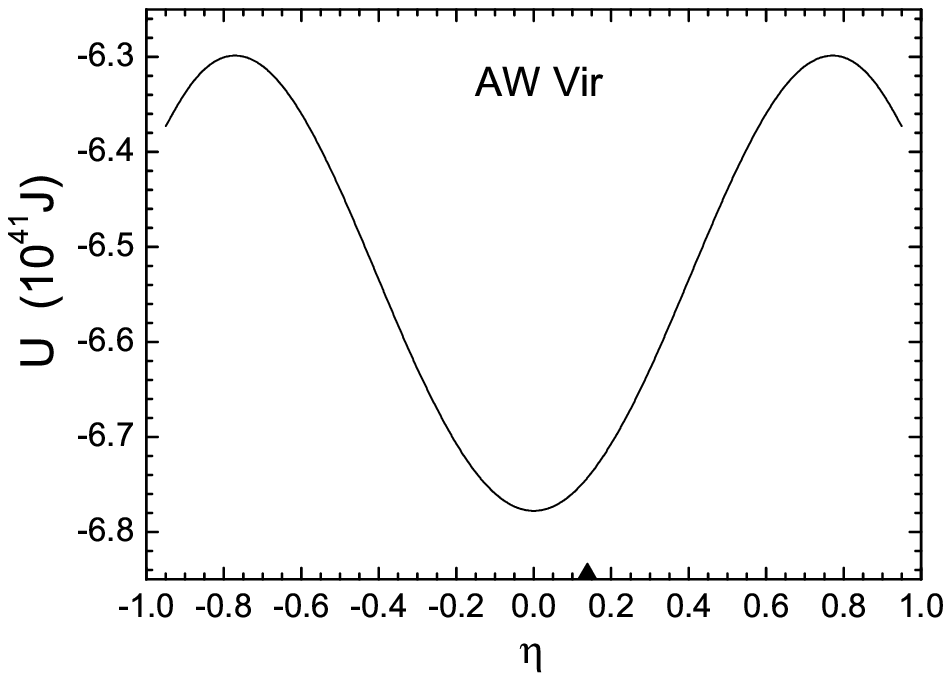}}
{\includegraphics[width=0.5\linewidth]{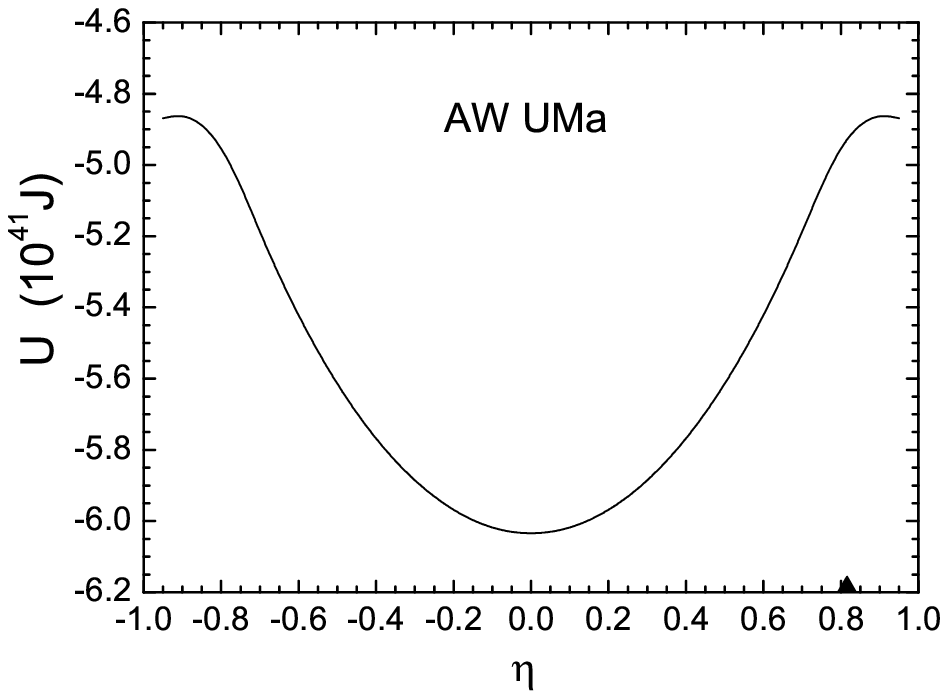}
\includegraphics[width=0.49\linewidth]{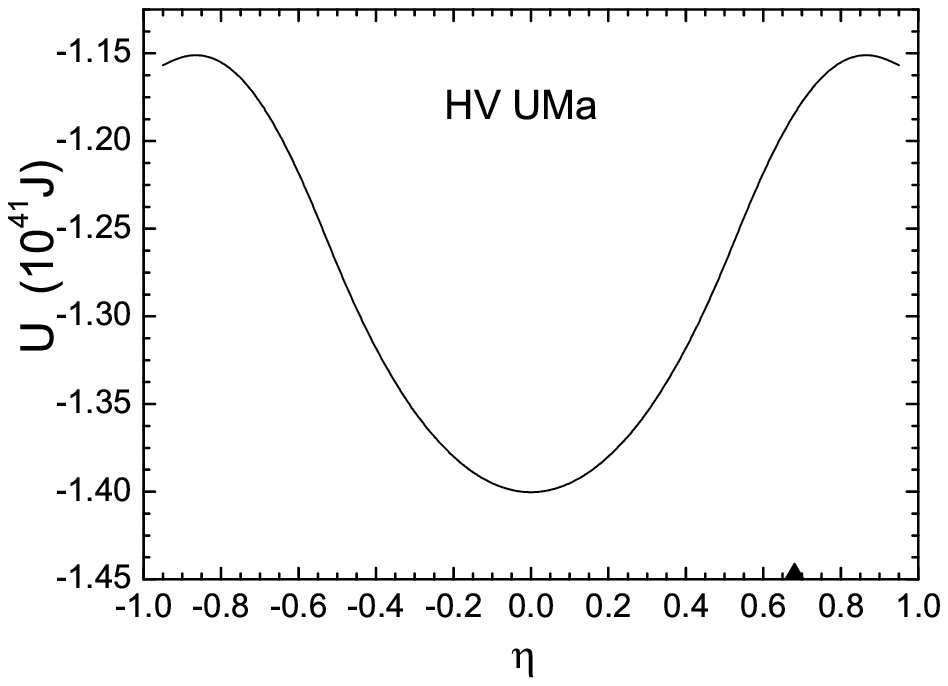}}
%
\caption{The calculated total potential energies $U$
vs $\eta$ for the indicated overcontact binary star systems. The arrow  on $x$-axis shows the corresponding
initial $\eta_i$ for binary star.
}
\label{1_fig}
\end{figure}

\begin{figure}[ht]
   \resizebox{\hsize}{!}
{\includegraphics[width=0.5\linewidth]{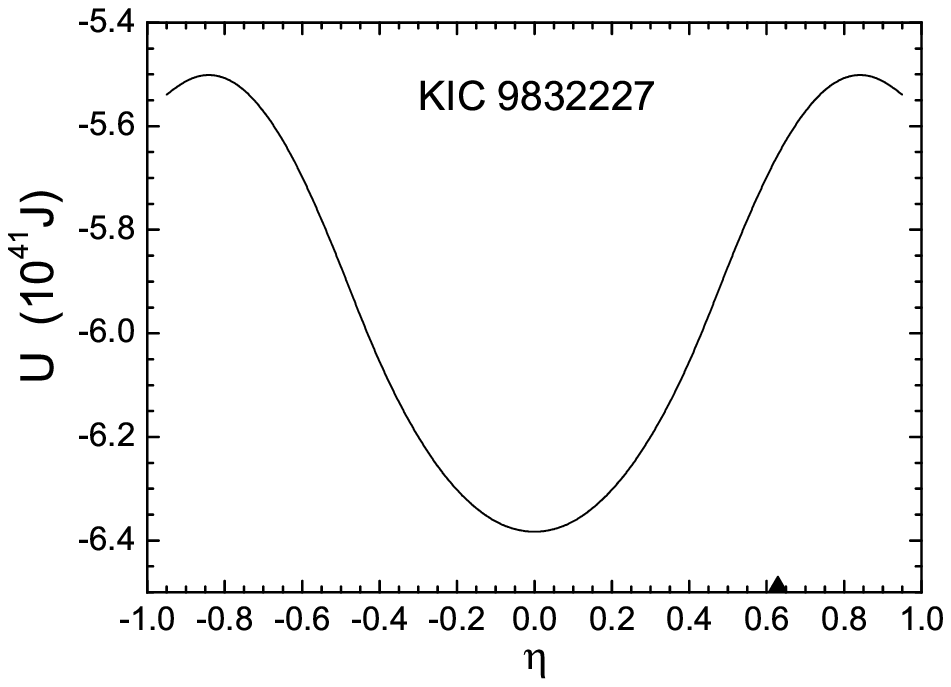}
\includegraphics[width=0.49\linewidth]{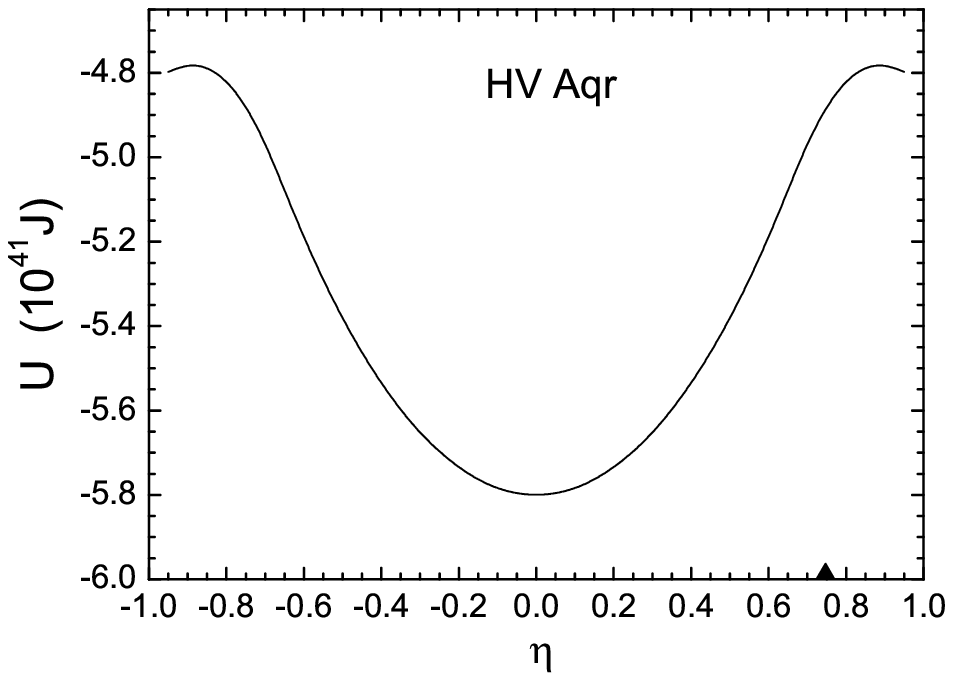}}
{\includegraphics[width=0.5\linewidth]{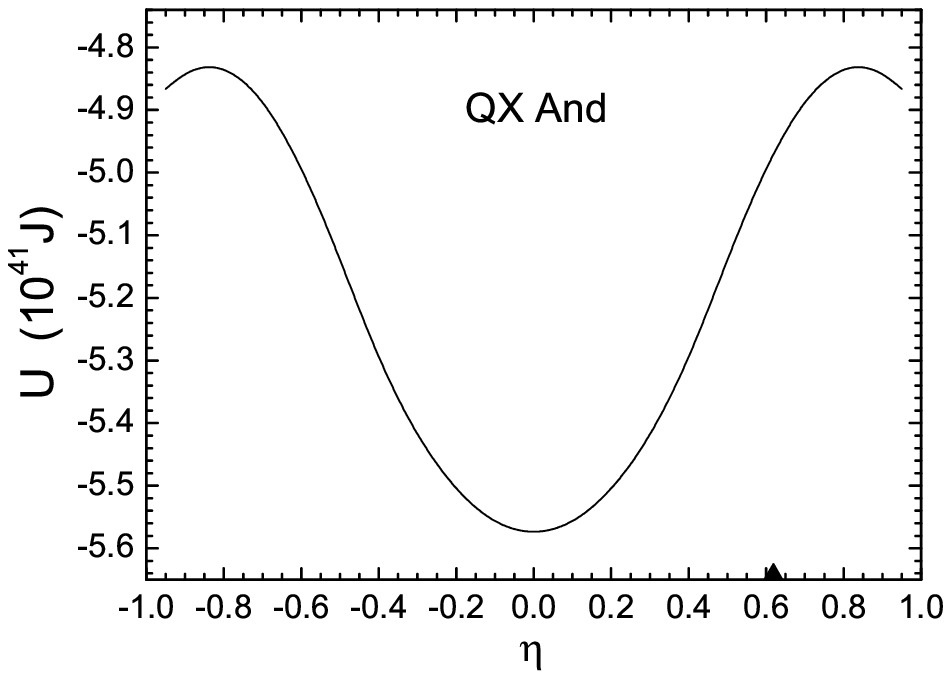}
\includegraphics[width=0.49\linewidth]{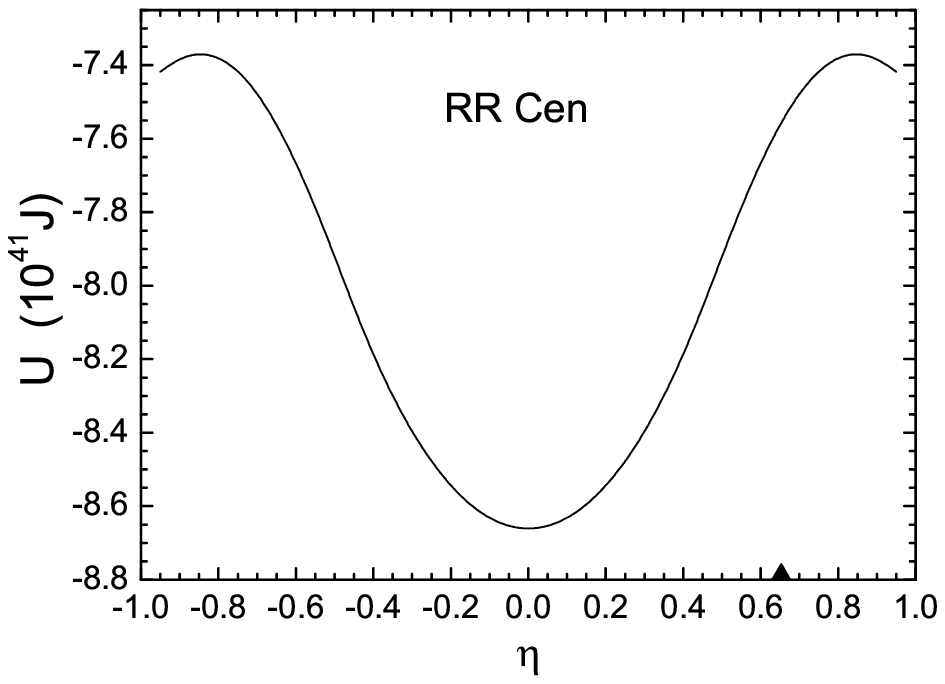}}
{\includegraphics[width=0.5\linewidth]{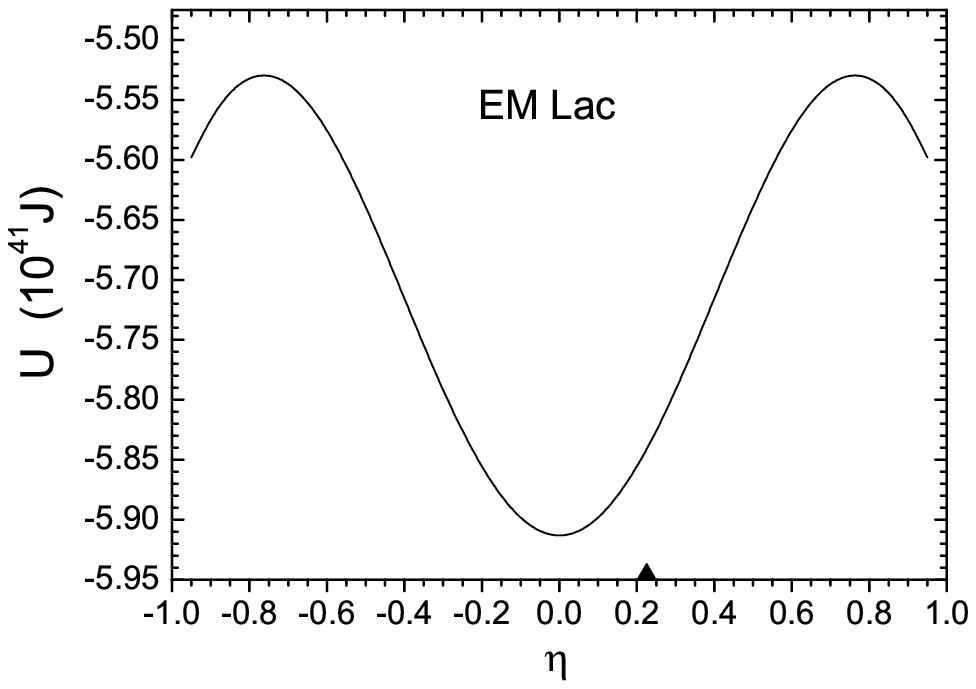}
\includegraphics[width=0.49\linewidth]{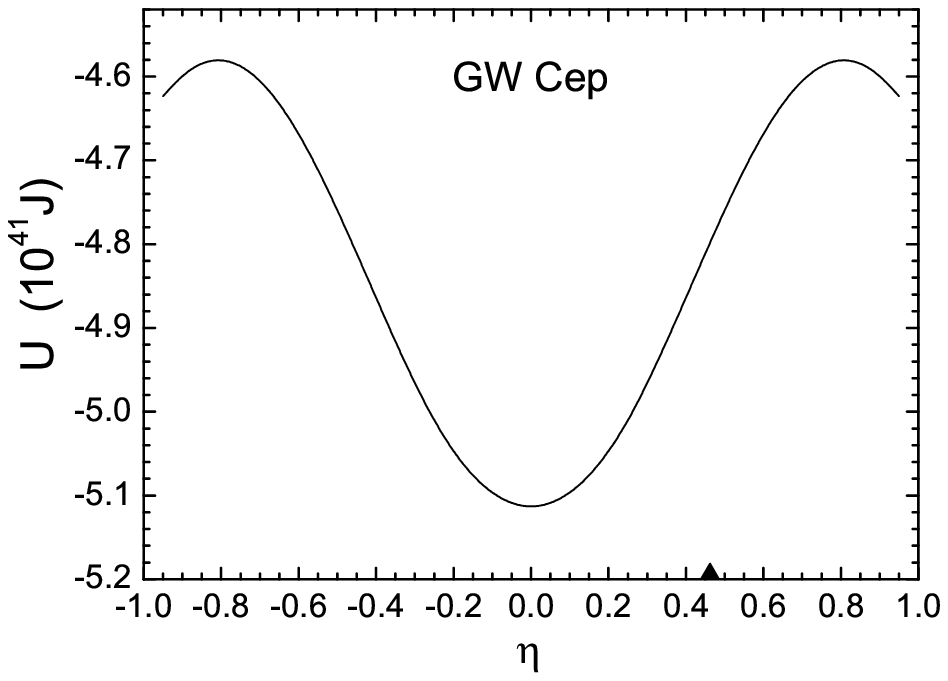}}
{\includegraphics[width=0.5\linewidth]{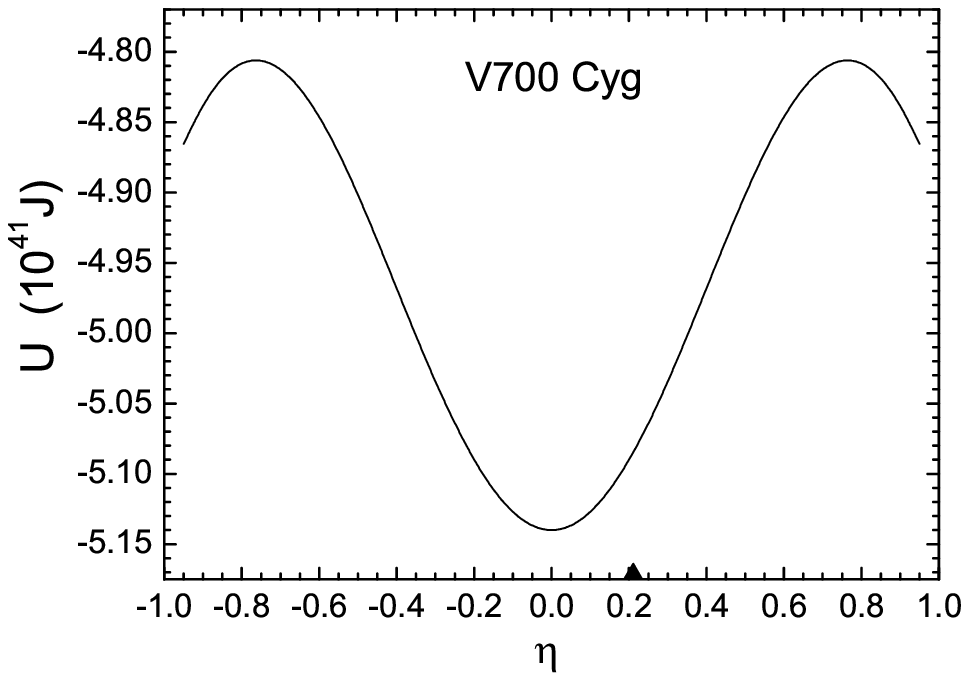}
\includegraphics[width=0.49\linewidth]{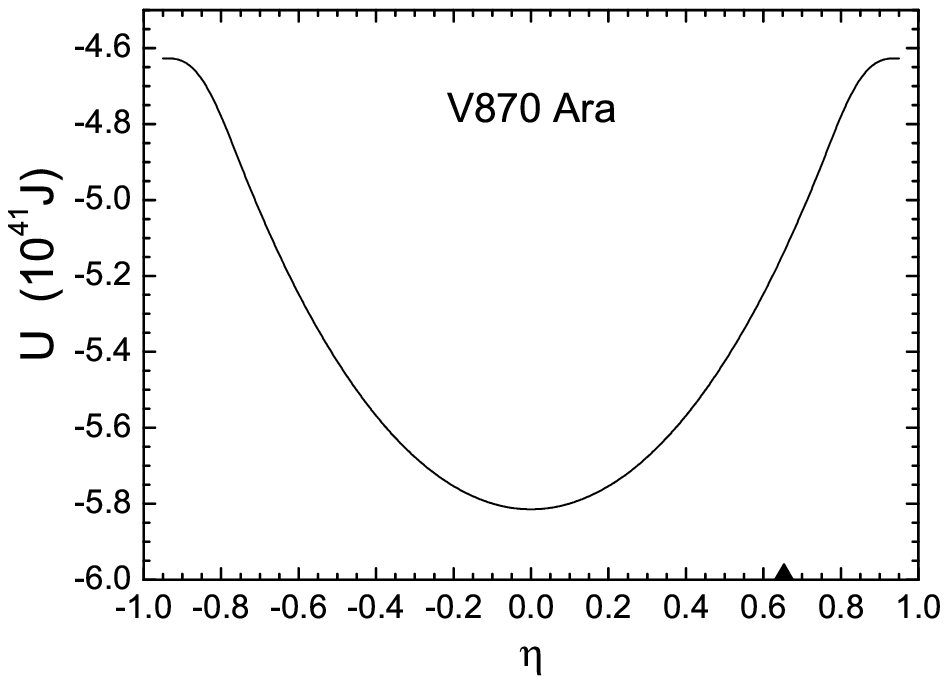}}
\caption{The same as in Fig. \ref{1_fig}, but
 for other indicated close binary star systems.
}
\label{2_fig}
\end{figure}

\begin{figure}[ht]
   \resizebox{\hsize}{!}
{\includegraphics[width=0.49\linewidth]{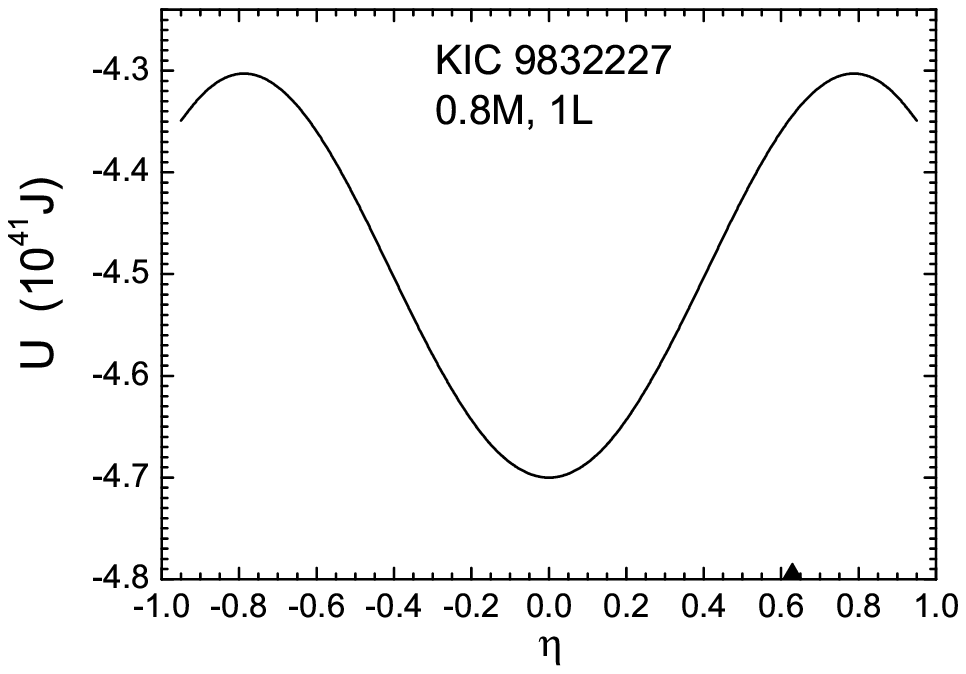}
\includegraphics[width=0.49\linewidth]{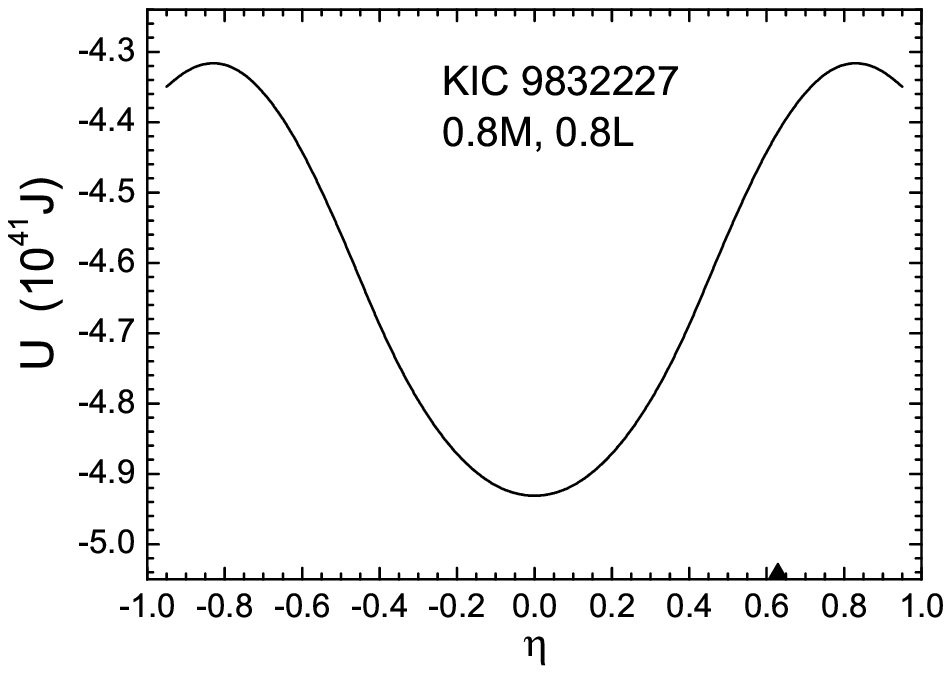}}
{\includegraphics[width=0.5\linewidth]{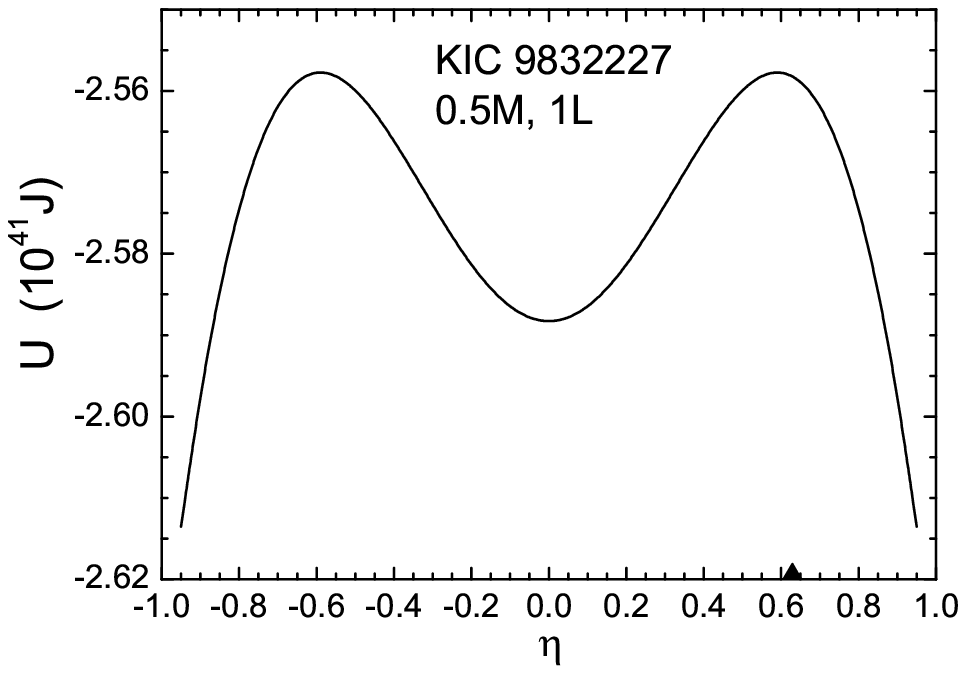}
\includegraphics[width=0.49\linewidth]{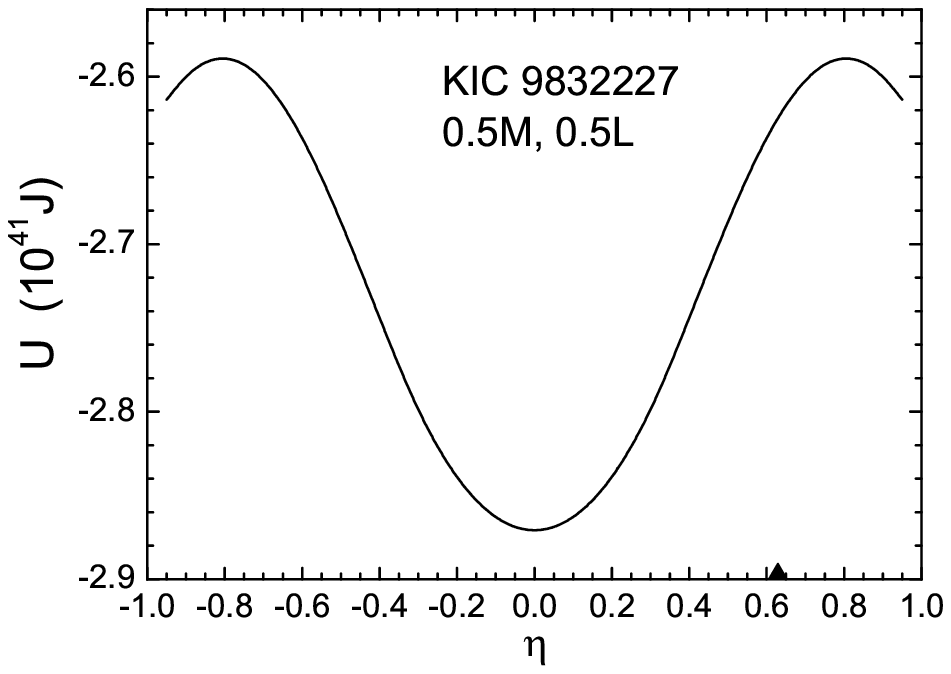}}
\caption{
The calculated total potential energies $U$, the relative  $R_m$ (solid line) and touching $R_t=R_1+R_2$ (dotted line)
distances between components in the units of the Sun radius
$R_{sun}=R_{\odot}$ vs $\eta$ for the   binary star KIC 9832227. The notations ($0.8M, 1L$), ($0.5M, 1L$), ($0.8M, 0.8L$), and ($0.5M, 0.5L$) mean
that the calculations are performed with the losses of   total mass $M$ and orbital angular momentum $L=L_i$ by  (20\%, 0\%),  (50\%, 0\%), (20\%, 20\%), and (50\%, 50\%),
respectively.
The arrow  on $x$-axis shows the corresponding
initial $\eta_i$ for binary star.
}
\label{3_fig}
\end{figure}

\begin{figure}[ht]
   \resizebox{\hsize}{!}
{\includegraphics[width=0.5\linewidth]{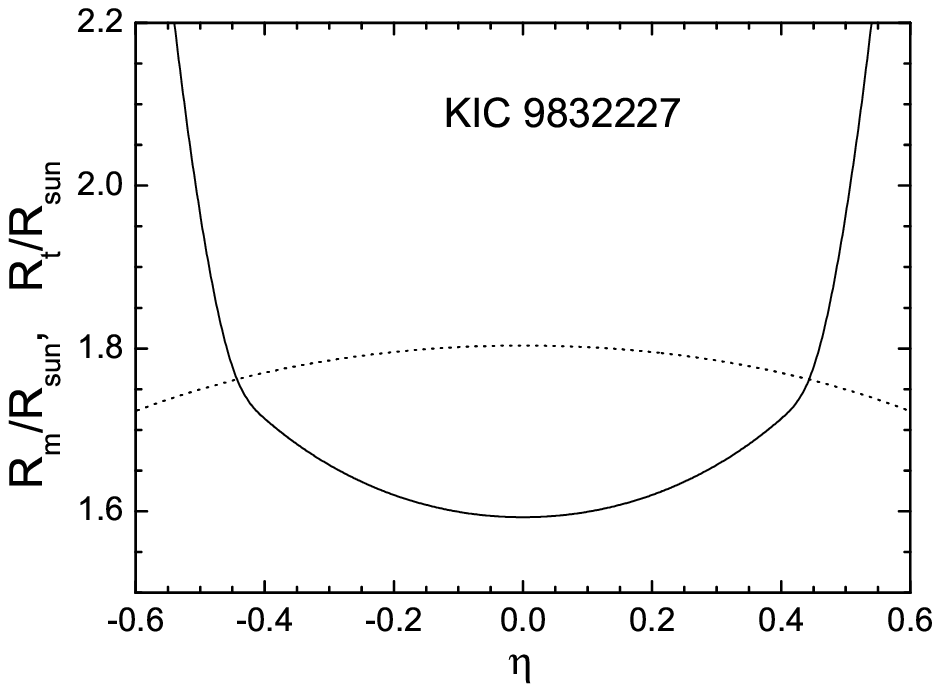}
\includegraphics[width=0.49\linewidth]{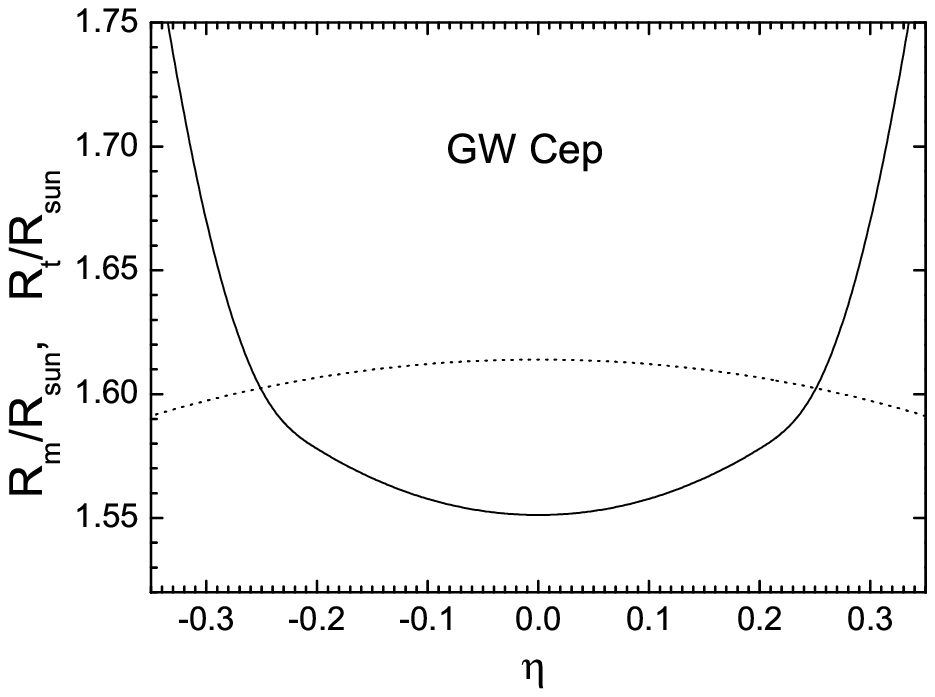}}
{\includegraphics[width=0.5\linewidth]{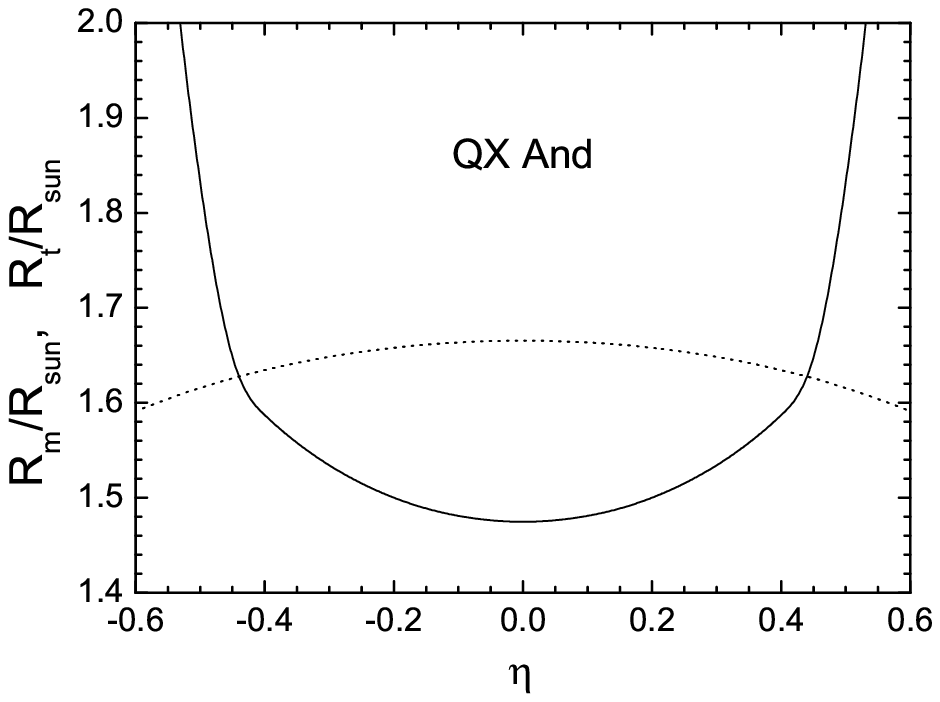}
\includegraphics[width=0.49\linewidth]{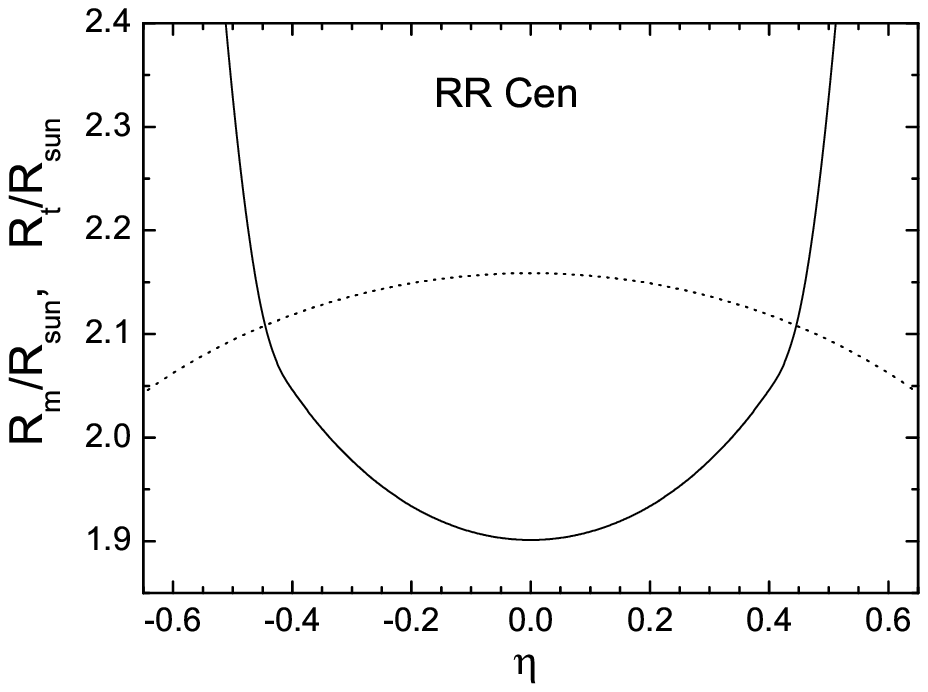}}
{\includegraphics[width=0.5\linewidth]{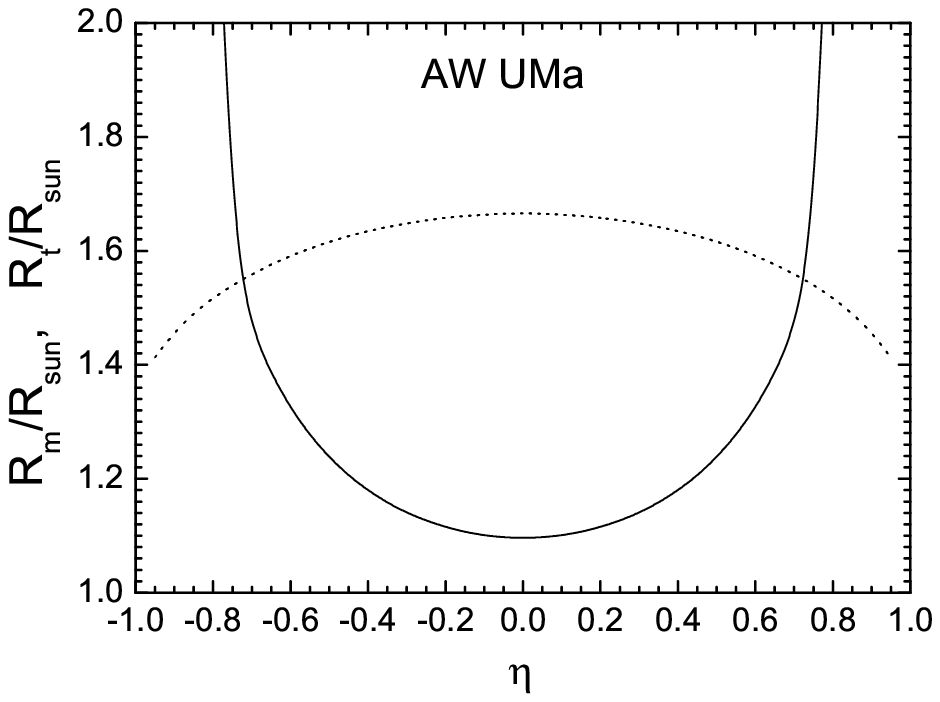}
\includegraphics[width=0.49\linewidth]{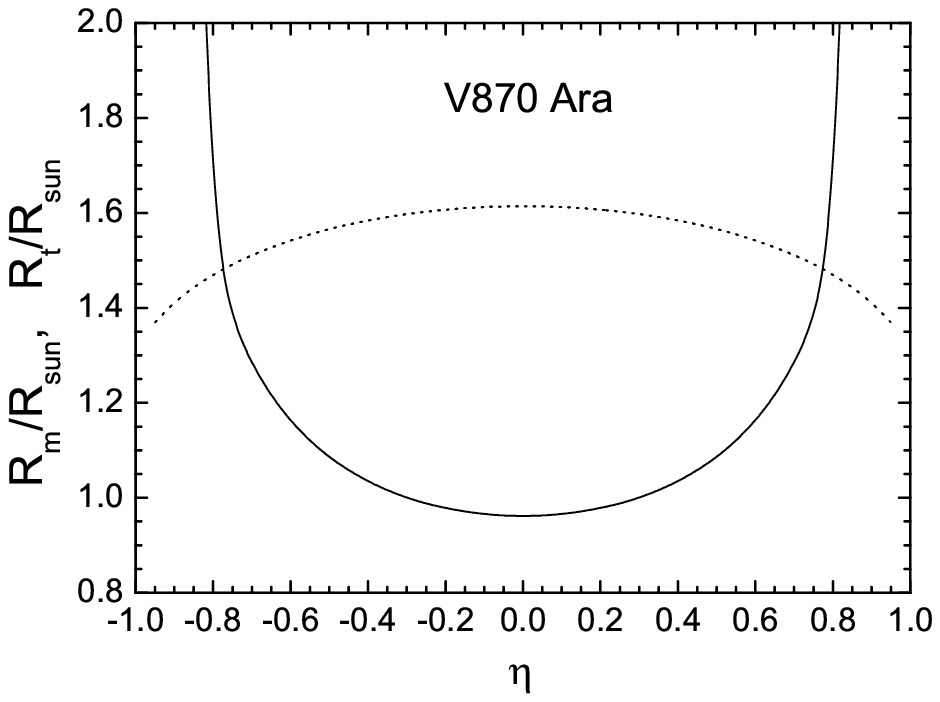}}
\caption{
The calculated  relative  $R_m$ (solid line) and touching $R_t=R_1+R_2$ (dotted line) distances between components
in the units of the Sun radius $R_{sun}=R_{\odot}$ vs $\eta$ for the indicated  binary star systems.
}
\label{4_fig}
\end{figure}


\end{document}